\documentclass[twocolumn,twoside,slac_two]{revtex4}
\usepackage{graphicx}
\usepackage{fancyhdr}
\input babarsym

\def\babar{\mbox{\slshape B\kern-0.1em{\smaller A}\kern-0.1em
    B\kern-0.1em{\smaller A\kern-0.2em R}}}
\def\piz   {\ensuremath{\pi^0}}
\def\Dbar  {\kern 0.2em\overline{\kern -0.2em D}{}}
\def\Dz    {\ensuremath{D^0}}
\def\Dzb   {\ensuremath{\Dbar^0}}
\def\to    {\ensuremath{\rightarrow}}
\newcommand{\Dzkkpz}{\ensuremath{\Dz \to K^{-}K^{+}\piz}}
\newcommand\fDz{\ensuremath{{f_{\Dz}}}}

\def\Dztilde   {\ensuremath {\tilde{D}^0}\xspace}

\def\Dzb   {\ensuremath {{\Db}^0}\xspace}
\def\Dztilde   {\ensuremath {\tilde{D}^0}\xspace}

\def \rb {\ensuremath {r_B}\xspace}

\def \deltab {\ensuremath {\delta_B}\xspace}

\def \xbmp {\ensuremath {x_\mp}\xspace}

\def \ybmp {\ensuremath {y_\mp}\xspace}

\newcommand{\re}{\ensuremath{\mathop{\rm Re}}}
\newcommand{\im}{\ensuremath{\mathop{\rm Im}}}

\def\bea{\begin{eqnarray}}
\def\eea{\end{eqnarray}}

\begin{document}

\title{Charm Dalitz Analyses}

%

\author{Gianluca Cavoto}
\affiliation{Universit\`a di Roma La Sapienza,  INFN and Dipartimento di Fisica, I-00185 Roma, Italy}

\begin{abstract}
 A review of recent  experimental results of  Dalitz analyses of charmed meson decays into three-body final states is presented. These analyses can help in understanding the strong interaction dynamics leading to the  observed light mesons spectrum (low mass scalar $\sigma$, $f_0(980)$, $a_0(980)$). A model for the decay amplitude into such states is very important for the extraction of the  angle $\gamma$ of the CKM unitarity triangle. Implications for such measurement are discussed.  
\end{abstract}

\maketitle

\thispagestyle{fancy}


\section{Introduction}

 A $D$ meson is  as a unique  "laboratory" to study light quark spectroscopy.
  It has a well defined spin-parity $J^P$ = $0^-$, constraining the angular momentum of the decay products in multibody final states which can be analyzed with the Dalitz plot technique \cite{dalitz}.
  
  Investigations of the low mass scalar mesons can be pursued 
   in three-body decays of pseudoscalar $D$ mesons giving their large coupling to such states.  The nature of such low mass scalar states is still under discussion  \cite{pennington}, since scalar mesons are difficult to resolve experimentally because of their large decay width. There are 
claims for the existence of broad states close to threshold such as $\kappa(800)$ and $\sigma(500)$~\cite{e791}.   On the theory side the scalar meson candidates are too numerous to fit in a single $J^{PC}=0^{++}$  $q \bar q$ nonet and therefore alternative interpretations are proposed. For instance, $a_0(980)$ or $f_0(980)$ may be 4-quark states due to their proximity to the $\bar K K$ threshold~\cite{q4}. 

These hypotheses can  be tested through 
an accurate measurement of branching fractions and couplings to  
different final states. In addition, comparison between the production
of these states in decays of differently flavored charmed 
mesons $D^0 (c \bar u)$, $D^+(c \bar d)$ and $D_s^+(c \bar s)$ \cite{chconj} 
can yield new information on their 
possible quark  composition. Another benefit of studying charm decays 
is that, in some cases, partial wave analyses
are able to isolate the scalar contribution 
almost background free.

Results of    $D^0$  Dalitz analyses can be an input for extracting the $C\!P$-violating phase  
$\gamma = \arg{\left(- V^{}_{ud} V_{ub}^\ast/ V^{}_{cd} V_{cb}^\ast\right)}$ 
of the quark mixing matrix by exploiting interference structure in the Dalitz 
plot from the decay $B^{\pm}\to D^0 K^{\pm}$~\cite{abi}.  Modeling of the 
$K \pi $ and $\pi \pi$ S-wave in $D$ decays is therefore an important element in such measurement, since the 
systematic uncertainty on $\gamma$  due to the Dalitz model is dominated by such components \cite{babar-kspp}.
 Model independent approaches using special Dalitz charm analyses are discussed and a projected systematic error on $\gamma$  in future experiments evaluated.

  \section{Dalitz analysis formalism}
   
   The amplitudes describing $D$ meson weak-decays into three-body final states 
are dominated by intermediate resonances that lead to highly non-uniform 
intensity distributions in the available phase space.

\indent Neglecting $C\!P$ violation in $D$ meson decays, we define the 
$D$ ($ \bar D$ ) decay amplitude $\cal{A}$ ($\bar{\cal{A}}$) 
in  a $D  \to    ABC$ Dalitz plot, as:
\begin{eqnarray}
\label{eq:1}
   {\cal{A}}[ D\to A B C ] \equiv \fDz(m_{B C}^2, m_{A C}^2),\\
\label{eq:2}
 \bar{\cal{A}}[ \bar D \to B  A C]  \equiv \fDz(m_{A C}^2, m_{B C}^2).
\end{eqnarray}

The complex quantum mechanical amplitude $f$ is a coherent sum of all relevant 
quasi-two-body $D\to(r\to AB)C$ resonances ("isobar model"~\cite{isobar}), 
$f = \sum_r a_r e^{i\phi_r} A_r(s)$. Here $s=m_{AB}^2$, and $A_r$ is the 
resonance amplitude. The  coefficients $a_r$ and $\phi_r$ are usually obtained  from 
a likelihood fit. The probability density function for the   signal events is $\left| f \right|^2$. Sub-modes branching fractions  ("fit fractions") are defined as 
$$f_r = \frac {|a_r|^2 \int |A_r|^2 dm_{AC}^2 dm_{BC}^2}
{\sum_{j,r} c_j c_r^* \int A_j A_r^* dm_{AC}^2 dm_{BC}^2}. $$
The fractions $f_r$ do not necessarily add up to 1 because of interference
effects among the amplitudes.
   
\indent For well established resonances  of the spin-1 (\textit{P-}wave) and spin-2 states,  the Breit-Wigner amplitude is used
\begin{eqnarray}
\label{eq:3}
A_{BW}(s) &=& {{\cal{M}}_L(s,p)}\hspace{2pt}{1\over{M_0^2-s-iM_0\Gamma(s)}},\\
\label{eq:4}
\Gamma(s) &=& \Gamma_0\Big({M_{0}\over \sqrt{s}}
\Big)\Big({p\over{p_{0}}}\Big)^{2L+1} {\Big[{{\cal{F}}_L(p) \over 
      {\cal{F}}_L(p_0)}\Big]^2},
\end{eqnarray}
\noindent where $M_0$ ($\Gamma_0$) is the resonance mass (width)~\cite{pdg}, 
$L$ is the angular momentum quantum number, $p$ is the momentum of either 
daughter in the resonance rest frame, and $p_0$ is the value of $p$ when s = 
$M_0^2$. The function ${\cal{F}}_L$ is the  Blatt-Weisskopf barrier 
factor~\cite{bw}: ${{{\cal{F}}_0}}$ = 1, ${{{\cal{F}}_1}}$ = 
$1/\sqrt{ 1+ {Rp}^2}$, and ${{{\cal{F}}_2}}$ = 
$1/\sqrt{ 9+ 3 {Rp}^2 + {Rp}^4}$, where we take the meson radial parameter $R$ is usually set to 
1.5 GeV${}^{-1}$~\cite{valR}. The spin part of the amplitude, 
${{\cal{M}}_L}$, is defined as: ${{\cal{M}}_0}$ = $M_{D}^2$, ${{\cal{M}}_1}$ = 
-2 $\vec{p_A}.\vec{p_C}$, and ${{\cal{M}}_2}$ = $4\over 3$ [
$3{(\vec{p_A}.\vec{p_C})}^2 - {|\vec{p_A}|}^2.{|\vec{p_C}|}^2$] 
$M_{D}^{-2}$, where $M_{D}$ is the nominal  $D$  mass, and $\vec{p_i}$ 
is the 3-momentum of particle $i$ in the resonance rest frame.\\

The $A_r(s)$ parameterization of the scalar   $f_0(980)$ resonance, 
whose mass, $m_{f_0}$, is close to the $K\overline{K}$ production threshold,
uses the Flatt\'e \cite{Flatte} formula
\begin{eqnarray}
\label{eqn:Flatte}
A_{f_0(980)}(m) =  \nonumber \\ 
\frac{1}{m_{f_0}^2 - s^2 - i [g^2_{f_0\pi\pi}        \rho_{\pi\pi}(s)
                                              + g^2_{f_0K\overline{K}} \rho_{K\overline{K}}(s)]},
\end{eqnarray}
where $g_{f_0\pi\pi}$ and $g_{f_0K\overline{K}}$ are the $f_0(980)$ coupling constants of
the resonance to the 
$\pi\pi$ and $K\overline{K}$ final states, and 
$\rho_{ab}(s) = 2p_a/\sqrt(s)$ is a phase space factor, calculated for the decay
products momentum, $p_a$, in the resonance rest frame. A similar formula is used for the $a_0(980)$
scalar resonance. 

 Different models for the  low mass $\pi\pi$ S wave,  (called $\sigma$ or $f_0(600)$ )  are used.
 In ~\cite{e791}  a simple spin-0 Breit-Wigner is tried.
Alternatively a  complex pole amplitude proposed in Ref.~\cite{Oller_2005} can be used
\begin{equation}
\label{eqn:ComplexPole}
A_{\sigma}(m) = \frac{1}{m^2_{\sigma} - m^2},
\end{equation}
where 
$m_{\sigma} = (0.47 - i 0.22)$~GeV is a pole position in the complex $s$ plane
estimated from the results of several experiments.
   
    More comprehensive parameterizations of the low mass $\pi\pi$ S wave has been proposed 
   and tested ~\cite{Schechter_2005}~\cite{Achasov_D3pi}.
    A K-matrix approach~\cite{ref:Kmatrix,ref:aitchison}, which gives a description of S wave $\pi\pi$ resonances treating the $\sigma$  and $f_0(980)$ contributions in a unified way has been used giving comparable results to the isobar technique~\cite{FOCUS_Dp-pipipi}.

   
 \section{ $\pi \pi$ S-wave.}

 \subsection{CLEO-c  $D^- \rightarrow  \pi ^+ \pi^- \pi^-$}

A study of charged $D$ decay to three charged pions 
has been carried out with the CLEO detector \cite{cleoc-Dp3pi}. This mode has been studied
previously by E687~\cite{E687_Dp-pipipi}, E691~\cite{E691_Dp-pipipi},
E791~\cite{e791}, and
FOCUS~\cite{FOCUS_Dp-pipipi}.  

E791 uses the isobar technique, where
each resonant contribution to the Dalitz plot
is modeled as a Breit-Wigner amplitude with a complex phase.
This works well for narrow, well separated resonances, but
when the resonances are wide and start to overlap,
solutions become ambiguous, and unitarity is violated.
In contrast, FOCUS uses the K-matrix approach.  The two techniques give a good description of
the observed Dalitz plots and agree about the
overall contributions of the resonances.
Both experiments see that about half of the fit fraction
for this decay is explained by a low $\pi^+\pi^-$ mass
S wave. 


The CLEO analysis  utilizes 281~pb$^{-1}$ of data collected on the $\psi(3770)$
resonance at $\sqrt{s}\simeq$3773~MeV at the Cornell Electron Storage Ring,
corresponding to a production of about $0.78\times 10^6$ $D^+D^-$ pairs. 
$D^+$ mesons are produced  close to the threshold, and are thus almost at rest.
Events from the decay $D^+ \to K^0_S\pi^+$, 
which has a large rate and contributes to the same final state, are isolated with the $\pi^+\pi^-$ invariant mass even without clearly detached vertexes as in 
the fixed target experiments.

 An isobar model is used to parametrize the signal decay  where the description
of the $\sigma$ from Ref.~\cite{Oller_2005} and the Flatt\'e
parameterization for the threshold effects on the
$f_0(980)$~\cite{Flatte} are included.  Alternative models are also tried and give comparably good fit results ~\cite{Schechter_2005}~\cite{Achasov_D3pi}.

The $D^+ \to \pi^-\pi^+\pi^+$ decay tracks are selected with requirements on their impact parameters  with respect to the beam spot. This removes $\sim$60\% of events with $K^0_S \to \pi^+\pi^-$ decays.
The remaining events from $D^+ \to K^0_S \pi^+$ represent about 
one third of those selected for the Dalitz plot.  Selection of events from the $D^+ \to \pi^-\pi^+\pi^+$ decay is done with two signal variables: $ \Delta E = E_D - E_{\mathrm{beam}} $ and $m_{\rm{BC}} = \sqrt{E^2_{\mathrm{beam}}-p^2_D}$, where $E_{\mathrm{beam}}$ is a beam energy, and
$E_D$ and $p_D$ are the energy and momentum of the reconstructed $D$ meson candidate, respectively. This gives 6991 events in the signal box, 2159$\pm$18 of these estimated to be background.

The presence of two $\pi^+$ mesons impose a Bose-symmetry
of the $\pi^-\pi^+\pi^+$ final state.
The Bose-symmetry when interchanging the two same sign charged pions
is explicitly accounted for in the  amplitude parameterization.
 Dalitz plot is analyzed  by choosing 
$x \equiv m^2(\pi^+\pi^-)_{\rm{Low}}$  and 
$y \equiv m^2(\pi^+\pi^-)_{\rm{High}}$ as the
independent ($x$,$y$) variables.  
The third variable $z \equiv m^2(\pi^+\pi^+)$ is dependent on $x$ and $y$
through the energy-momentum balance equation.

In the  Dalitz plot analysis events in the band
$0.2 < m^2(\pi^+\pi^-)_{\rm{Low}} < 0.3$~(GeV/$c^2$)$^2$ are excluded
which is approximately ten times our $K^0_S \to \pi^+\pi^-$ mass
resolution. This leaves for the  Dalitz plot analysis  4086 events
  $\sim$2600 of which are  signal events.

CLEOc was able to  reproduce the fit results E791~\cite{e791}.  The amplitude normalization and sign conventions are different from E791, in particular the inclusion of a $\sigma \pi$ contribution  gives  a fit probability of $\simeq 20\%$. Possible contributions form all  known $\pi^+\pi^-$ resonances listed in Ref.~\cite{PDG_2004} were tried, including high mass resonances giving  asymptotic ``tails" at the edge of the  kinematically allowed region.

For the $f_0(980)$  the Flatt\'{e} formula, Eq.~\ref{eqn:Flatte}, is used
with parameters taken from the recent BES~II measurement \cite{BES_2005}.
For the $\sigma$   a complex pole amplitude, Eq.~\ref{eqn:ComplexPole}, was eventually tried
rather than the spin-0 Breit-Wigner.

\begin{table}[!htb]
\caption{Results of the isobar model analysis of the $D^+\to\pi^-\pi^+\pi^+$ Dalitz plot.  
         For each contribution the relative amplitude, phase, and fit fraction
         is given.  The errors are statistical and systematic, respectively.}
\begin{tabular}{|c|c|c|c|}
\hline
\hline
Mode                &  Amplitude (a.u.)      & Phase ($^\circ$)  & Fit fraction (\%)     \\
\hline		     
$\rho(770)\pi^+$    &  1(fixed)             &  0(fixed)          &  20.0$\pm$2.3$\pm$0.9 \\
$f_0(980)\pi^+$     &  1.4$\pm$0.2$\pm$0.2  &   12$\pm$10$\pm$5  &  4.1$\pm$0.9$\pm$0.3  \\
$f_2(1270)\pi^+$    &  2.1$\pm$0.2$\pm$0.1  & --123$\pm$6$\pm$3  &  18.2$\pm$2.6$\pm$0.7 \\
$f_0(1370)\pi^+$    &  1.3$\pm$0.4$\pm$0.2  & --21$\pm$15$\pm$14 &  2.6$\pm$1.8$\pm$0.6  \\
$f_0(1500)\pi^+$    &  1.1$\pm$0.3$\pm$0.2  & --44$\pm$13$\pm$16 &  3.4$\pm$1.0$\pm$0.8  \\
$\sigma$ pole       &  3.7$\pm$0.3$\pm$0.2  &  --3$\pm$4$\pm$2   &  41.8$\pm$1.4$\pm$2.5 \\
\hline		     
\hline
\end{tabular} 
\label{tab:Dp-3pi_results} 
\end{table}

\begin{figure}[!ht]
\begin{center}
\begin{tabular} {cc}  
{\includegraphics[height=4.5cm]{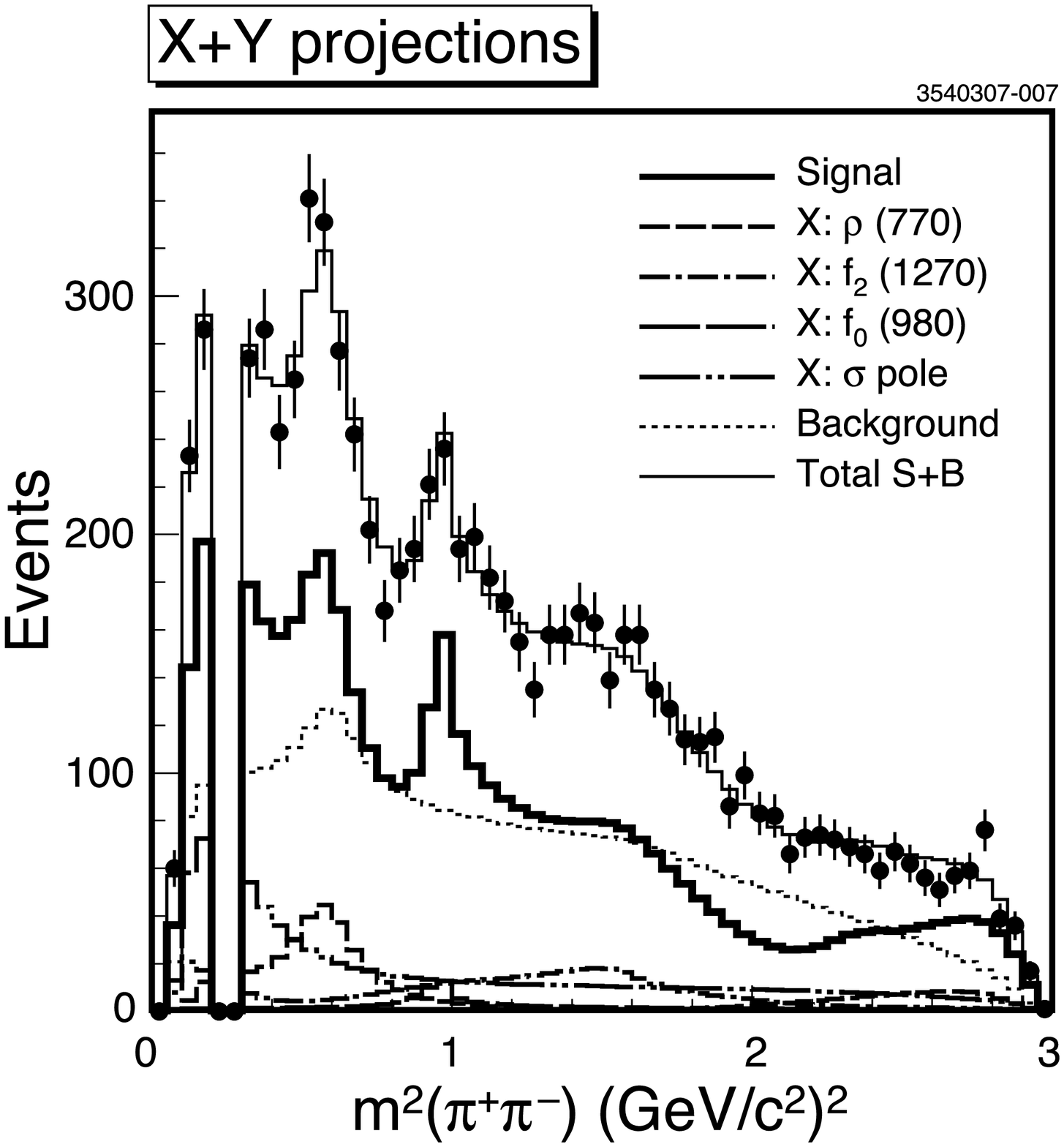}} &
{\includegraphics[height=4.5cm]{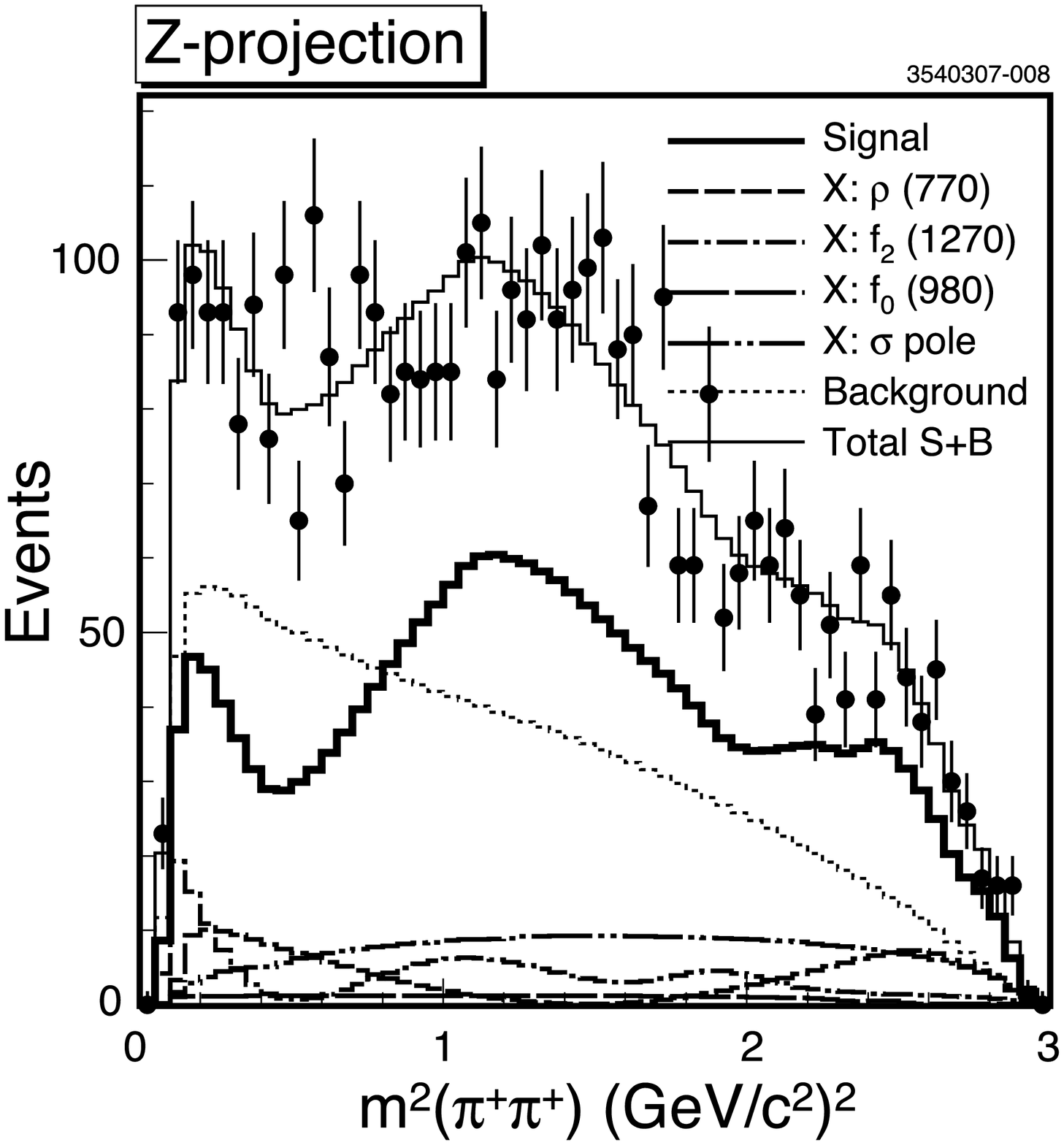}} 
\end{tabular}   
\caption{ Projection of the Dalitz plot  onto the $m^2(\pi^+\pi^-)$  axis 
             (two combinations per $D^+$ candidate) for CLEO-c data (points)
             and isobar model fit (histograms) showing the various components (left).Projection of the Dalitz plot onto the $m^2(\pi^+\pi^+)$ axis (right).
              }
\label{fig:cleoC-dalitz}
\end{center}
\end{figure}

Table~\ref{tab:Dp-3pi_results} shows the list of  surviving contributions with their fitted amplitudes and phases, and calculated fit fractions after a procedure of addition and removal of resonances to improve the consistency  between the model and data. 
The sum of all fit fractions is 90.1\%, and 
the fit probability is $\simeq$28\% for 90 degrees of freedom.
The two projections of the Dalitz plot and selected fit components are shown in Fig.~\ref{fig:cleoC-dalitz}

For contributions that are not significant upper limits at the  95\% confidence level are set. 

The systematic uncertainties, shown in Table~\ref{tab:Dp-3pi_results}, 
are estimated from numerous fit variations, by adding or removing degrees of freedom, changing the event selection, and varying the
efficiency and background parameterizations.

 For the poorly established resonances
 as the $\sigma$ pole, their parameters  are allowed to float and the variations
of the other fit parameters contribute to the systematic
errors. The  fitted values for the $\sigma$ pole  are $Re(m_\sigma)$ (MeV/$c^2$)  = 466$\pm$18  and 
 $Im(m_\sigma)$ (MeV/$c^2$)  = --223$\pm$28.

 \subsection{BaBar  $D_s  \rightarrow   K^+  K^-  \pi^+$}

\begin{figure}[!ht]
\begin{center}
 \includegraphics[height=7.5cm]{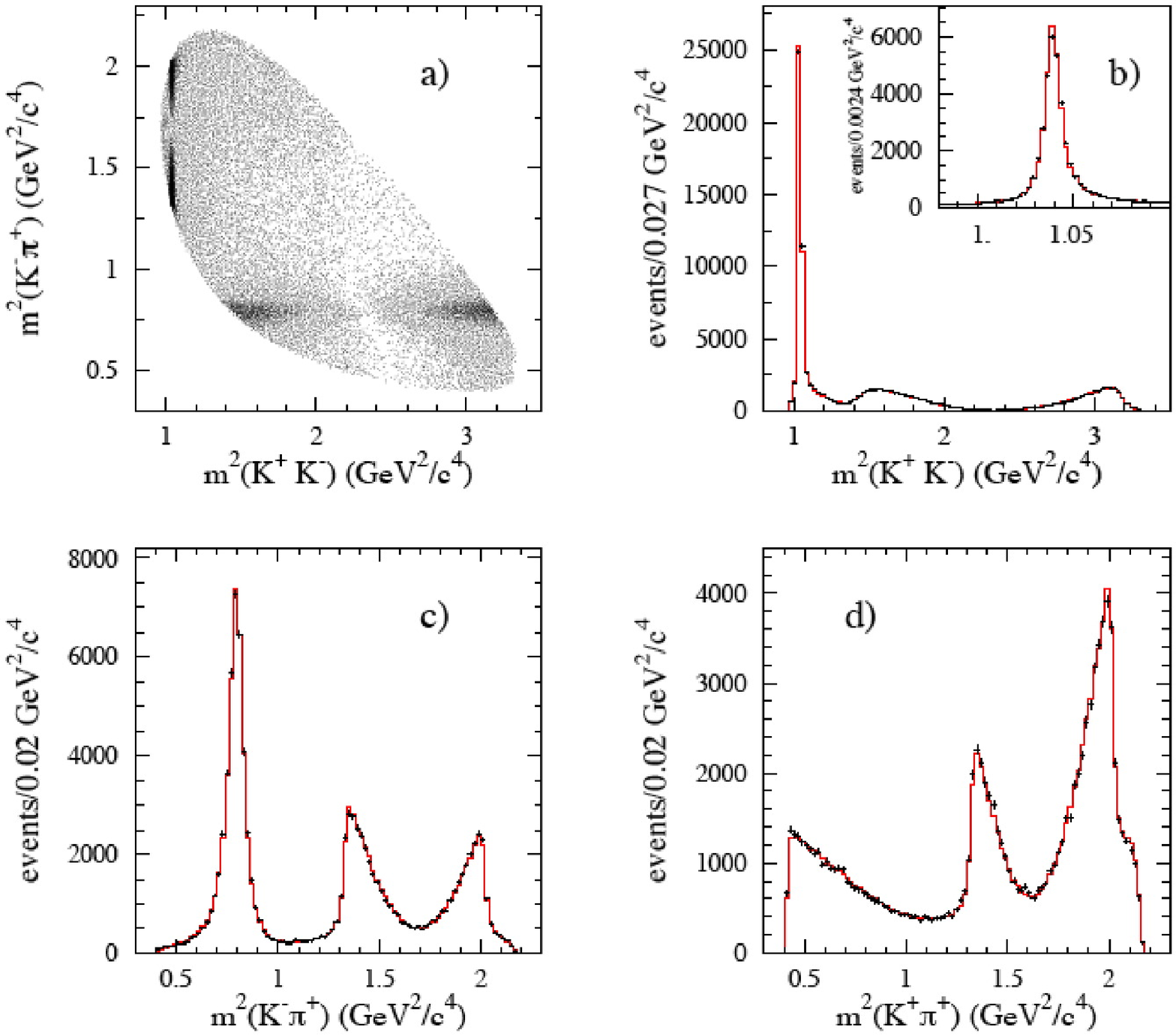}
\caption{ Dalitz plot of $D_s \to  \pi^+  K^+ K^-$  }
\label{fig:babar-Ds-kkpi-dalitz}
\end{center}
\end{figure}

 BaBar analyzed 240 fb$^{-1}$ taken at the center of mass energies near the $\Upsilon (4S)$ resonance. Events are selected in a sample of events having at least three reconstructed  charged tracks with two well identified
   kaons and one pion. The three tracks are fit to a common vertex with the constraint they come from the beamspot. The decay chain $D_s^*(2112)^+ \to D^+_s \gamma $  helps in discriminating signal from combinatorial background.  Additional requirements based on  kinematic and geometric information are combined to further suppress the background. The final sample contains 100850 events with a purity of 95\%. An unbinned maximum likelihood fit of the Dalitz plot  (Fig.\ref{fig:babar-Ds-kkpi-dalitz}) is performed to extract the relative amplitudes and phases of the intermediate resonances as shown in Tab.~\ref{tab:DsKKpi_results}. The decay is dominated by the $\phi(1020)\pi^+$ and $f_0(980)\pi^+$. The  $f_0(980)$ is parametrized with a coupled channel Breit-Wigner  \cite{BES_2005} and its  contribution  is large but it is subject to a large systematic error due to the poor knowledge of its parameters and possible $a_0(980)$ contributions that are difficult to disentangle in the $K \bar K$ projection. Analysis of the angular moment distribution confirms such picture with a big S-wave--P-wave interference in the $K \bar K$ channel in the region of the  $\phi(1020)$. On the other hand very small activity is present in the  $ K^*(892) $ region suggesting a small $K \pi$ S-wave, and therefore no evidence of a $\kappa (800)$.
   
\begin{table*}[!htb]
\caption{Results of the isobar model analysis of the $ D_s  \to  K^+  K^-  \pi^+$ Dalitz plot.  
         For each contribution the relative amplitude, phase, and fit fraction
         is given.  The errors are statistical and systematic, respectively.}
\begin{tabular}{|c|c|c|c|}
\hline
\hline

Mode                &  Amplitude (a.u.)      & Phase ($^\circ$)  & Fit fraction (\%)     \\
\hline		     

 $ K^*(892)  K^+$    &   $1$(fixed)                & $  0$(fixed)                                          & $  48.7 \pm 0.2 \pm 1.6  $\\
$\phi(1020)\pi^+$       & $   1.081 \pm 0.006 \pm 0.049     $ & $   2.56 \pm 0.02 \pm 0.38    $ & $  37.9 \pm 0.2 \pm 1.8 $  \\
$f_0(980)\pi^+$      & $    4.6 \pm 0.1 \pm 1.6                       $ & $ -1.04 \pm 0.04 \pm 0.48        $ & $   35 \pm 1  \pm 14 $\\
$K^*_0(1430)^0  K^+$    & $  1.07 \pm 0.06 \pm 0.73      $ & $  -1.37 \pm 0.05 \pm 0.81  $ & $ 2.0 \pm 0.2 \pm 3.3  $\\
$f_0(1710)\pi^+$      & $  0.83 \pm 0.02 \pm 0.18                $ & $  -2.11 \pm 0.05 \pm 0.42   $ & $   2.0 \pm 0.1 \pm 1.0  $\\
$f_0(1370)\pi^+$      & $   1.74 \pm 0.09 \pm 1.05               $ & $  -2.6 \pm 0.1 \pm 1.1   $ & $    6.3 \pm 0.6 \pm 4.8  $\\
$K^*_2(1430)^0  K^+$     & $   0.43 \pm 0.05 \pm 0.34     $ & $  -2.5 \pm 0.1 \pm 0.3   $ & $  0.17 \pm 0.05 \pm 0.30 $ \\
$f_2(1270)\pi^+$      & $   0.40 \pm 0.04 \pm 0.35              $ & $  0.3 \pm 0.2 \pm 0.5  $ & $    0.18 \pm 0.03 \pm 0.40  $\\
Sum          & $    132 \pm 1  \pm 16  $ &    &     \\
\hline		     
\hline
\end{tabular} 
\label{tab:DsKKpi_results} 
\end{table*}


 \subsection{BaBar  $D^0 \rightarrow   \bar K^0  K^-  K^+$}

   The data sample used in the BaBar  $D^0 \rightarrow   \bar K^0  K^-  K^+$  analysis consists of 91.5 \invfb recorded  with the \babar\ detector at the SLAC \pep2\ storage rings \cite{babar-Dkkk0} . 
The PEP-II facility operates nominally at the \Y4S 
resonance, 
providing collisions of 9.0 \gev electrons on 3.1 \gev positrons. The data 
set includes 
82~\invfb collected in this configuration
(on-resonance) and 9.6~\invfb collected at a c.m. energy 40 MeV below
the \Y4S resonance (off-resonance).

\begin{figure}
\begin{center}
\includegraphics[width=7cm]{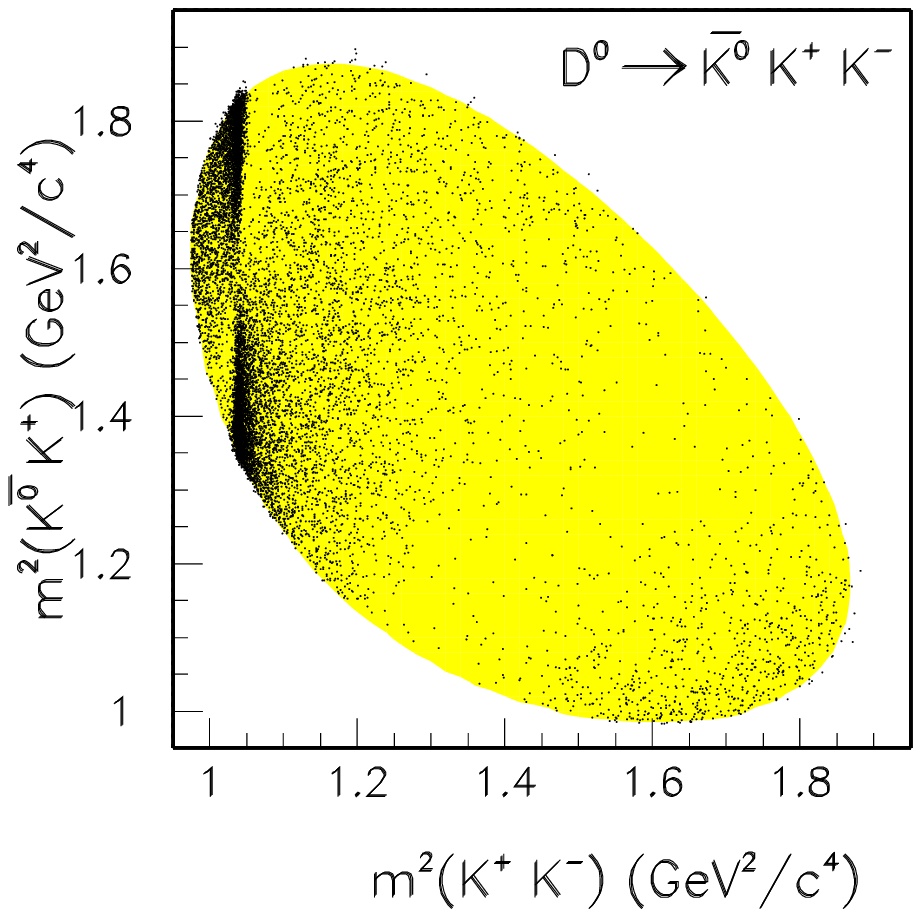}
\caption{Dalitz plot of $D^0 \to \overline{K}{}^0 K^+ K^-$.
} 
\label{fig:fig3}
\end{center}
\end{figure}

Selecting events within $\pm 2 \sigma$ of the fitted $D^0$ mass value,
a signal fraction of 97.3\% is obtained for the 12540 events selected. 
The Dalitz plot for these $D^0 \to \overline{K}{}^0 K^+ K^-$ candidates 
is shown 
in Fig.~\ref{fig:fig3}.
In the $K^+ K^-$ threshold region, a strong $\phi(1020)$ signal is observed, 
together with a rather broad structure. 
A large asymmetry with respect to the $\overline{K}{}^0 K^+$ axis can also 
be seen 
in the vicinity of the $\phi(1020)$ signal, which is most probably the 
result of interference between  
$S$ and $P$-wave amplitude contributions to the $K^+ K^-$ system.
The $f_0(980)$ and $a_0(980)$ $S$-wave resonances are, in fact, just below the 
$K^+ K^-$ 
threshold, and might be expected to contribute in the vicinity of $\phi(1020)$.
An accumulation of 
events due to a charged $a_0(980)^+$ can be observed on the lower right edge 
of the Dalitz plot. This contribution, however, does not overlap with 
the $\phi(1020)$ region and 
this allows the $K^+K^-$ scalar and vector components to be separated using a 
partial wave analysis in the low mass $K^+K^-$ region.

The helicity angle, $\theta_K$, is then defined as the 
angle between the $K^+$ for $D^0$ (or $K^-$ for $\overline{D}{}^0$)
in the $K^+ K^-$ rest frame and 
the $K^+ K^-$ direction in the $D^0$ (or $\overline{K}{}^0$)
rest frame. The $K^+ K^-$ mass distribution has been modified by weighting 
each $D^0$ candidate by the spherical harmonic $Y_L^0(\cos \theta_K)$ (L=0-4)
divided by its (Dalitz-plot-dependent) fitted efficiency. It is found that all the $\left<Y^0_L \right>$ moments are
small or consistent with zero, except for 
$\left<Y^0_0 \right>$, $\left<Y^0_1 \right>$ and $\left<Y^0_2 \right>$.
  
In order to interpret these distributions a simple 
partial wave analysis has been performed, involving only $S$- and
$P$-wave amplitudes. This results in the following set of equations~\cite{chung}:

$$\sqrt{4 \pi} \left<Y^0_0 \right> = S^2 + P^2$$
$$\sqrt{4 \pi} \left<Y^0_1 \right> = 2 \mid S \mid \mid P \mid \cos \phi_{SP} \qquad (3)$$
$$\sqrt{4 \pi} \left<Y^0_2 \right> = \frac{2}{\sqrt 5} P^2,$$

\noindent
where $S$ and $P$ are proportional to the size of the $S$- and $P$-wave
contributions and $\phi_{SP}$ is their relative phase.
Under these assumptions, the $\left<Y^0_2 \right>$ moment is proportional to $P^2$ 
so that it is natural that the $\phi(1020)$ appears free of background, 
as is observed.
This distribution
has been fit using the following relativistic $P$-wave Breit-Wigner, yielding  the
following parameters:
\begin{center}
$m_\phi$ = 1019.63 $\pm$ 0.07, $\Gamma_\phi$ = 4.28 $\pm$ 0.13 MeV/$c^2$
\end{center}
in agreement with PDG values (statistical errors only).

\begin{figure}[!ht]
\begin{center}
\includegraphics[height=8.0cm]{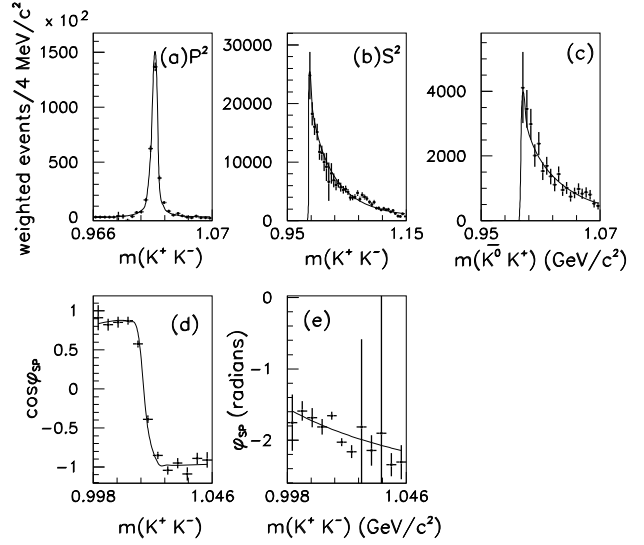} 
\caption{ Results from the $K^+ K^-$ Partial Wave Analysis corrected 
for phase space. (a) $P$-wave strength, (b) $S$-wave strength.
(c) $m(\overline{K}{}^0 K^+)$ distribution, (d) $\cos\phi_{SP}$
in the $\phi(1020)$ region. (e) $\phi_{SP}$ in the threshold
region after having subtracted the fitted $\phi(1020)$ phase motion 
shown in (d). The lines correspond to the fit described in the text. }
\label{fig:fig7}
\end{center}
\end{figure}

The above system of equations  can be solved directly for $S^2$, $P^2$ and
$\cos \phi_{SP}$ and corrected for phase space distribution. 
The phase space corrected spectra are shown in Fig.~\ref{fig:fig7}.

The distributions have been fitted using a model with  $\phi(1020)$ for the P-wave ,  a scalar contribution in the $K^+ K^-$ mass projection  entirely due to the $a_0(980)^0$, $\overline{K}{}^0 K^+$ mass distribution is entirely due to  $a_0(980)^+$  and the $\cos \phi_{SP}$  described with BW models.

The $a_0(980)$ scalar resonance has a mass very close
to the $\bar K K$ threshold and decays mostly to $\eta \pi$. It has been
described by a coupled channel Breit Wigner of the form:
$$BW_{ch}(a_0)(m) = \frac {g_{\bar K K}}{m^2_0 - m^2 - 
i(\rho_{\eta \pi} g_{\eta \pi}^2 + \rho_{\bar K K} g_{\bar K K}^2)} \quad (5)$$
\noindent
where $\rho(m) = 2 q/m$ while $g_{\eta \pi}$ and $g_{\bar K K}$ describe
the $a_0(980)$ couplings to the $\eta \pi$ and $\bar K K$ systems respectively.

The best measurements of the $a_0(980)$ parameters come from 
the Crystal Barrel experiment~\cite{cbar}, 
in $\bar p p$ annihilations, with  a value of $g_{\bar K K}=329 \pm 27$ (MeV)$^{1/2}$.
  $m_0$ and $g_{\eta \pi}$ have been fixed to the 
Crystal Barrel measurements, but   $g_{\bar K K}$, on the other hand, has been fit  (stat only) $g_{\bar K K}$ = 464 $\pm$ 29 (MeV)$^{1/2}.$

 The determination of $g_{\bar K K}$ has been redone  
in a complete Dalitz plot analysis with an evaluation  of the 
 systematic error. The fit produces a reasonable representation of the data for all of the 
projections. The $\chi^2$ computed on the Dalitz plot gives a value of 
$\chi^2/NDF$=983/774. The sum of the fractions is $130.7 \pm 2.2 \pm 8.4$\%.
The regions of higher $\chi^2$ are distributed rather uniformly on the Dalitz
plot. Attempts to improve the fit quality by including other  scalar amplitudes caused the 
fit to diverge, producing a sum of fractions well above 200\% along with small
improvements of the fit quality.

The final fit results showing fractions, amplitudes and phases
are summarised in Table~\ref{tab:res}. 
For $\overline{K}{}^0 f_0(980)$ and $K^+ a_0(980)^-$ (DCS), being
consistent
with zero,  only the fractions have 
been tabulated. 
For the Dalitz plot analysis  the $f_0(980)$ contribution is  found to be consistent with zero,

\begin{table*}[tbp]
\caption{Results from the Dalitz plot analysis of $D^0 \to \overline{K}{}^0 K^+ K^-$. The fits have been performed using the value of $g_{\bar K K} = 464$ (MeV)$^{1/2}$ resulting from the partial wave analysis.}
\label{tab:res}
\begin{center}
\vskip -0.2cm
\begin{tabular}{lllll}
\hline
Final state & Amplitude & Phase (radians) & Fraction (\%) \cr
\hline
$\overline{K}{}^0 a_0(980)^0$ & 1. & 0. & 66.4 $\pm$ 1.6 $\pm$ 7.0 \cr
\hline
$\overline{K}{}^0 \phi(1020)$ & 0.437 $\pm$ 0.006 $\pm$ 0.060 & 1.91 $\pm$ 0.02 $\pm$
 0.10 &  45.9 $\pm$ 0.7 $\pm$ 0.7 \cr
\hline 
$K^- a_0(980)^+$ & 0.460 $\pm$ 0.017 $\pm$ 0.056 & 3.59 $\pm$ 0.05 $\pm$ 0.20 &
 13.4 $\pm$ 1.1 $\pm$ 3.7 \cr
\hline 
$\overline{K}{}^0 f_0(1400)$ & 0.435 $\pm$ 0.033 $\pm$ 0.162 & -2.63 $\pm$ 0.10 $\pm$ 0
.71 & 3.8 $\pm$ 0.7 $\pm$ 2.3 \cr
\hline
$\overline{K}{}^0 f_0(980)$ & & &  0.4 $\pm$ 0.2 $\pm$ 0.8 \cr 
\hline 
$K^+ a_0(980)^-$ & & &  0.8 $\pm$ 0.3 $\pm$ 0.8 \cr
\hline
Sum & & & 130.7 $\pm$ 2.2 $\pm$ 8.4 \cr
\hline
\end{tabular}
\end{center}
\end{table*} 

A test has been performed by leaving $g_{\bar K K}$ as a free parameter 
in the Dalitz plot analysis, the resulting central value of $g_{\bar K K}$  being 
$$g_{\bar K K} = 473 \pm 29 \;(\text{stat.}) \pm 40 \;(\text{syst.}) ({\rm MeV})^{1/2}.$$
This value differs significantly from the Crystal Barrel measurement. An improvement 
of this measurement can be foreseen by adding data from the $a_0(980) \to \eta \pi$ decay 
mode such as $D^0 \to K^0_s \eta \pi^0$. 

It must be noticed that   reliable estimate of the expected contribution of the
$f_0(980)$ in $D^0 \to \overline{K}{}^0 K^+ K^-$ decay is not possible until more accurate
measurements of the $f_0(980)$ parameters and couplings become available.
This can be performed, for example, by using high 
statistics samples of $D^+_s \to \bar K K \pi^+$ and $D^+_s \to \pi^+ \pi^+ \pi^-$ decays.

 \section{$K \pi $  S-wave }
   
   The $K^{\pm}\piz$ systems from the decay \Dzkkpz ~can provide information on the $K\pi$ \textit{S-}wave 
(spin-0) amplitude in the mass range 0.6--1.4 \gevcc, and hence on the 
possible existence of the $\kappa(800)$, reported to date only in the neutral 
state ($\kappa^0 \to K^- \pi^+$)~\cite{e791}. If the $\kappa$ has isospin 
$1/2$, it should be observable also in the charged states. 


   \subsection{BaBar $D^0 \to K^+ K^- \pi^0$}

\begin{figure}[!htbp]
\begin{center}
\begin{tabular}{cc} 
\includegraphics[width=0.235\textwidth]{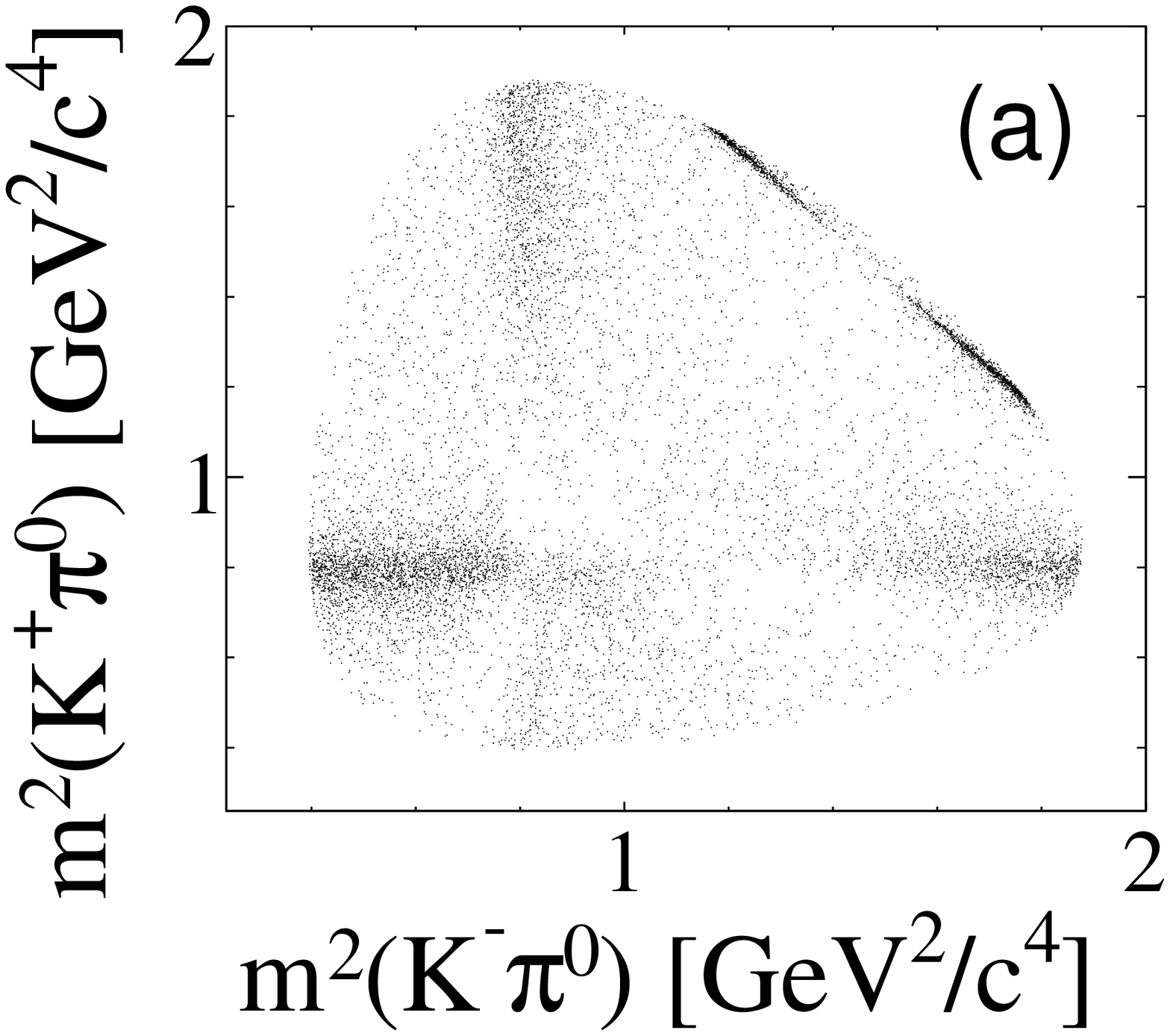}
&
\includegraphics[width=0.235\textwidth]{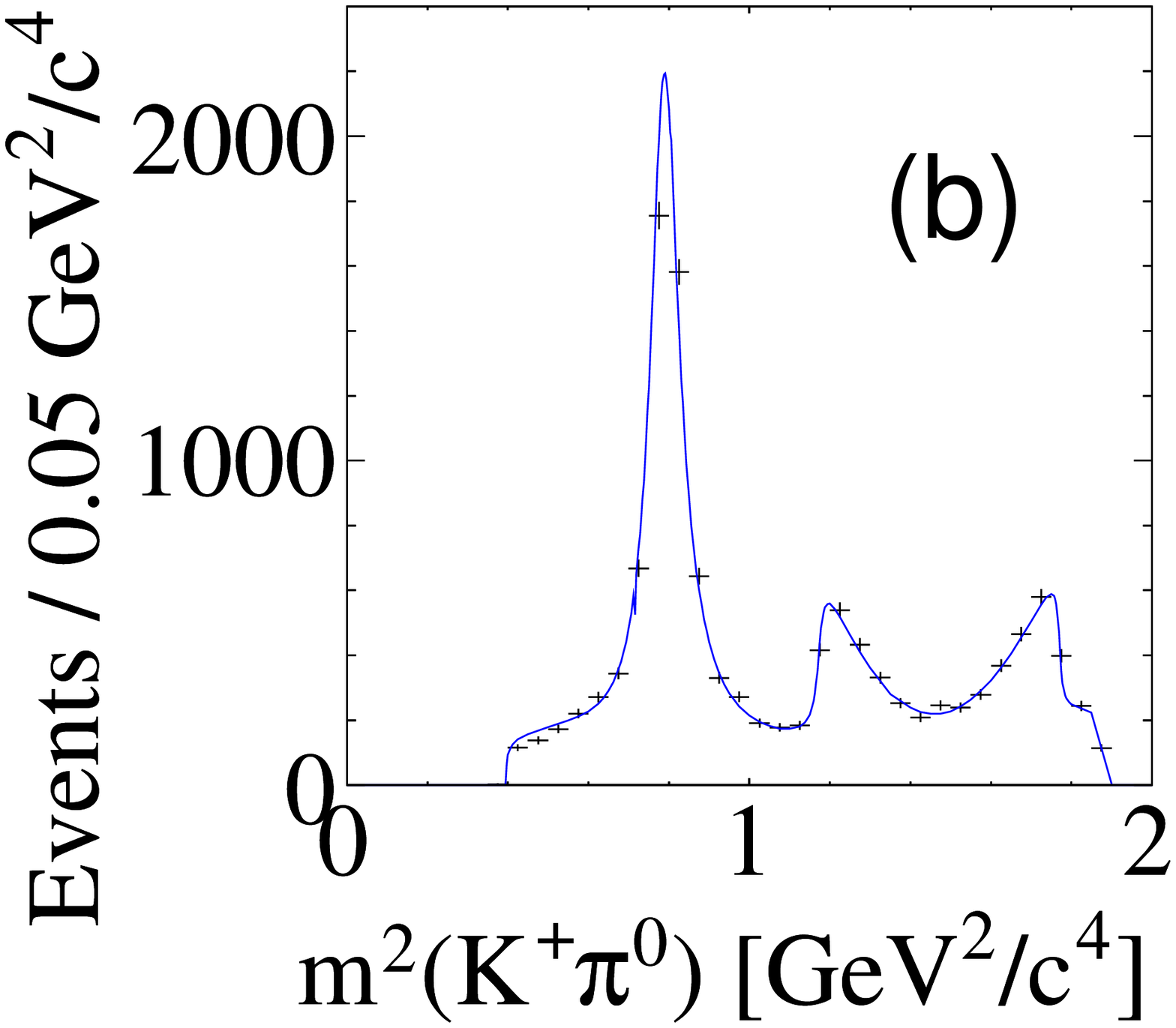}\\
\end{tabular}
\begin{tabular}{cc} 
\includegraphics[width=0.235\textwidth]{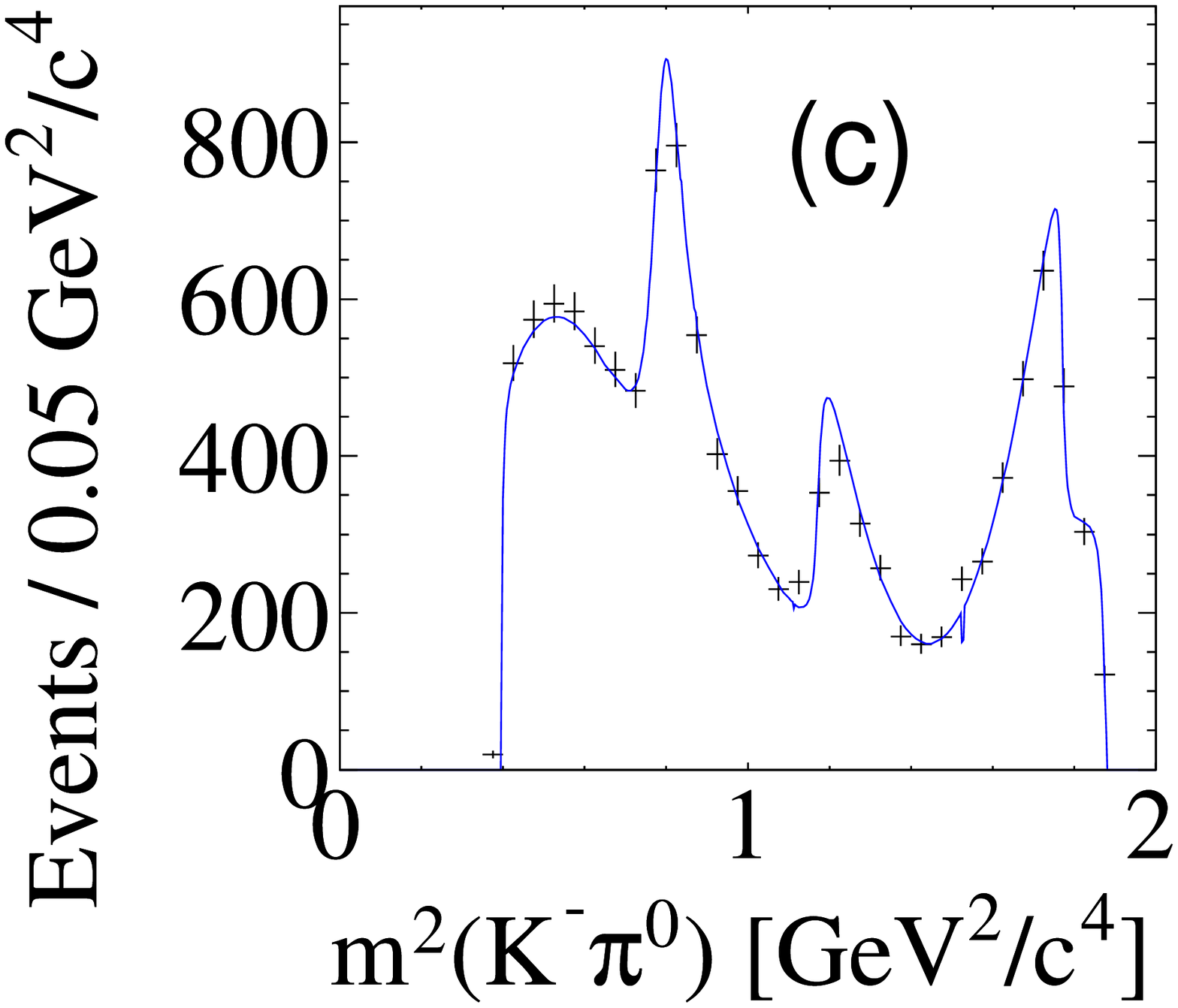}
&
\includegraphics[width=0.235\textwidth]{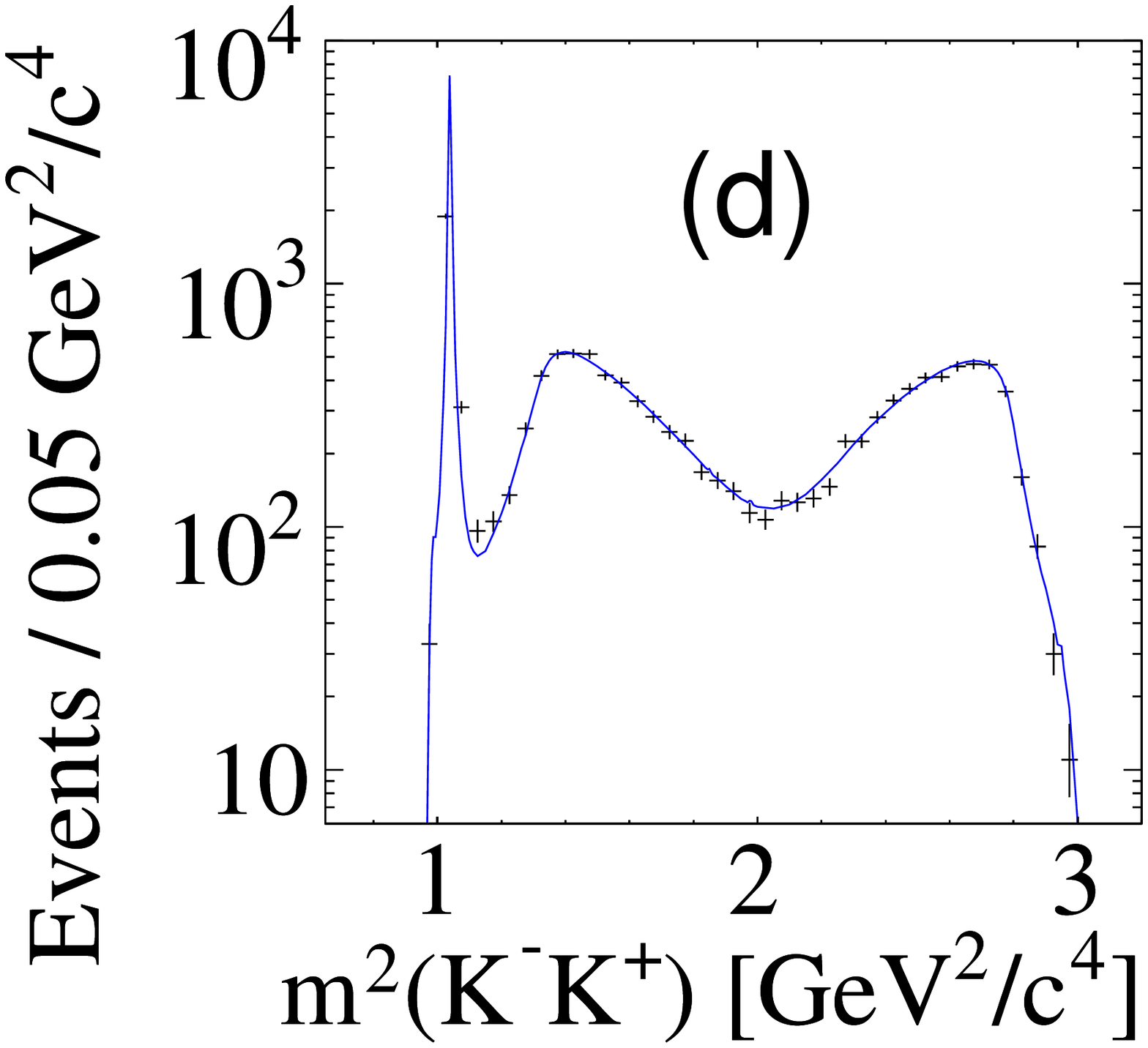}\\
\end{tabular}
\vspace{-1em}
\caption{(Color online) Dalitz plot for $D^0 \to K^- K^+ \piz$ 
data (a), and the corresponding squared invariant mass projections (b--d). 
The three-body invariant mass of the \Dz\ candidate is constrained to the 
nominal value. In plots~(b--d), the dots (with error bars, black) are data 
points and the solid lines (blue) correspond to the best isobar fit models.}
\label{babar-KKpi0}
\end{center}
\end{figure}

\begin{table*}[htbp]
\caption{The results obtained from the $D^0 \to K^- K^+ \piz$ Dalitz plot 
fit. We define amplitude coefficients, $a_r$ and $\phi_r$, relative to those 
of the $K^{*}(892)^+$. The errors are statistical and systematic, 
respectively. We show the $a_0(980)$ contribution, when it is included in 
place of the $f_0(980)$, in square brackets. We denote the 
$K\pi$ \textit{S-}wave states here by $K^{\pm}\piz(S)$.}
\label{tab:result}
\begin{tabular}{|l|rrr|}
\hline\hline
State &   Amplitude, $a_r$ &  Phase, $\phi_r$ (${}^\circ$) & Fraction, 
 $f_r$ (\%) \cr 
 \hline\hline
$K^*(892)^{+}$ & 1.0 (fixed) & 0.0 (fixed) & 45.2$\pm$0.8$\pm$0.6 \cr
$K^*(1410)^{+}$ &  2.29$\pm$0.37$\pm$0.20  &   86.7$\pm$12.0$\pm$9.6  
&  3.7$\pm$1.1$\pm$1.1 \cr
$K^+\piz(\textit{S})$  &  1.76$\pm$0.36$\pm$0.18 & -179.8$\pm$21.3$\pm$12.3 
&  16.3$\pm$3.4$\pm$2.1\cr
$\phi(1020)$ &  0.69$\pm$0.01$\pm$0.02  &  -20.7$\pm$13.6$\pm$9.3  
&  19.3$\pm$0.6$\pm$0.4 \cr
$f_0(980)$ &  0.51$\pm$0.07$\pm$0.04  & -177.5$\pm$13.7$\pm$8.6  
&  6.7$\pm$1.4$\pm$1.2 \cr
$\left[a_0(980)^0\right]$ & [0.48$\pm$0.08$\pm$0.04]& [-154.0$\pm$14.1$\pm$8.6]
&  [6.0$\pm$1.8$\pm$1.2] \cr
$f_2'(1525)$  &  1.11$\pm$0.38$\pm$0.28  &  -18.7$\pm$19.3$\pm$13.6 &  
0.08$\pm$0.04$\pm$0.05  \cr
$K^*(892)^{-}$ &  0.601$\pm$0.011$\pm$0.011 &  -37.0$\pm$1.9$\pm$2.2   
&  16.0$\pm$0.8$\pm$0.6 \cr
$K^*(1410)^{-}$ &  2.63$\pm$0.51$\pm$0.47  & -172.0$\pm$6.6$\pm$6.2   
&  4.8$\pm$1.8$\pm$1.2  \cr
$K^-\piz(\textit{S})$& 0.70$\pm$0.27$\pm$0.24 & 133.2$\pm$22.5$\pm$25.2 
& 2.7$\pm$1.4$\pm$0.8 \cr
\hline 
\end{tabular}
\end{table*}

 BaBar analyzed  385 fb${}^{-1}$ of $e^+e^-$ 
collision data and  reconstructed 
the decays $D^{*+}\to D^0 \pi^{+}$  with $D^0 \to K^- K^+ \piz$ \cite{babar-kkpi0}. Requirements on   the center-of-mass  momentum of the $D^0$ candidate and on $|m_{D^{*+}} - m_{D^0}|$ yields   in the signal region, 
$1855 < m_{D^0} < 1875$~\mevcc  $11278 \pm 110$ signal events with 
a purity of about 98.1\%.

\indent For \Dz\ decays to $K^{\pm}\piz$ \textit{S-}wave states,  three 
amplitude models have been considered. One model uses the LASS amplitude for  
$K^-\pi^+\to K^-\pi^+$ elastic scattering~\cite{LASS},
\begin{eqnarray}
  \label{eq:5}
  A_{{K\pi}(S)} (s) = {\sqrt{s} \over p}\sin\delta(s) e^{i\delta(s)},\indent{}
\indent{}\indent{}\indent{}\indent{}\indent{}\indent{}\indent{}\indent{}\\
  \label{eq:6}
  \delta(s) = \cot^{-1}\Big({1\over pa} + {bp\over 2}\Big) 
              + \cot^{-1}\Big({M_{0}^2-s\over M_{0}\Gamma_{0}\cdot{M_{0}\over 
		  \sqrt{s}}\cdot{p\over{p_{0}}}}\Big),
\end{eqnarray}
\noindent where $M_{0}$ ($\Gamma_{0}$) refers to the $K^*_0(1430)$ mass 
(width), $a =1.95 \pm 0.09$ GeV${}^{-1}c$, and $b = 1.76 \pm 0.36$ 
GeV${}^{-1}c$. The unitary nature of Eq.~\ref{eq:5} provides a good 
description of the amplitude up to 1.45 \gevcc\ (i.e., $K\eta^{\prime}$ 
threshold). In Eq.~\ref{eq:6}, the first term is a non-resonant contribution 
defined by a scattering length $a$ and an effective range $b$, and the 
second term represents the $K^*_0(1430)$ resonance. The phase space factor 
$\sqrt{s}/ p$ converts the scattering amplitude to the invariant amplitude. 

A second model uses the E791 results for the $K^-\pi^+$ \textit{S-}wave 
amplitude from an energy-independent partial wave analysis in the 
decay $D^+\to K^-\pi^+\pi^+$~\cite{brian}. The third model uses a coherent 
sum of a uniform non-resonant term, and Breit-Wigner terms for the 
$\kappa(800)$ and $K^*_0(1430)$ resonances.\\

\indent The \Dz decay to a $K^-K^+$ \textit{S-}wave state is described by a 
coupled-channel Breit-Wigner amplitude for the $f_0(980)$ and $a_0(980)$ 
resonances, with their respective couplings to $\pi\pi$, $K \bar{K}$ and 
$\eta\pi$, $K\bar{K}$ final states~\cite{Flatte},
\begin{equation}
A_{f_0[a_0]} (s) = \frac{M_{\Dz}^2} {M_{0}^2 - s - 
  i (g_1^2~\rho_{\pi\pi\left[\eta\pi\right]} + g_2^2~\rho_{K\bar{K}})}.
\label{eq:7}
\end{equation}

\indent  Several models are used  incorporating  
various combinations of intermediate states. 
In each fit,  the $K^*(892)^{+}$ is included  and the complex amplitude 
coefficients of other states relative to it is measured.
 
\indent The LASS $K\pi$ \textit{S-}wave amplitude gives the best agreement 
with data and it is uses it in the nominal fits. 
The $K\pi$ \textit{S-}wave modeled by the combination of $\kappa(800)$ (with 
parameters taken from Ref.~\cite{kappa}), a non-resonant term and 
$K^*_0(1430)$ has a smaller fit probability ($\chi^2$ probability $<$ 5\%). 
The best fit with this model ($\chi^2$ probability 13\%) yields a charged 
$\kappa$ of mass (870 $\pm$ 30)~\mevcc, and width (150 $\pm$ 20)~\mevcc, 
significantly different from those reported in Ref.~\cite{kappa} for the 
neutral state. This does not support the hypothesis that production of a 
charged, scalar $\kappa$ is being observed. The E791 amplitude~\cite{brian} 
describes the data well, except near threshold ($\chi^2$ probability  23\%). 
 Analysis of  moments of $\cos{\theta_H}$ confirms little variation in S-wave phase up to about 1.02-1.03 GeV/c$^2$ and matched the behaviour obtained with the isobar model. 

The  results of the best fit  are summarized 
in Table~\ref{tab:result}.   Neglecting $C\!P$ violation, the strong phase difference, $\delta_D$, 
between the \Dzb and \Dz decays to $K^*(892)^{+}K^-$ state and their amplitude 
ratio, $r_D$, are given by
\begin{equation}
r_D e^{i\delta_D} = \frac{a_{\Dz\to K^{*-}K^+}}{a_{\Dz\to K^{*+}K^-}} 
{ } e^{i(\delta_{K^{*-}K^+}{ } - { } \delta_{K^{*+}K^-})}. 
\end{equation}
Combining the results of models I and II, we find $\delta_D$ = 
$-35.5^\circ \pm 1.9^\circ$ (stat) $\pm 2.2^\circ$ (syst) and $r_D$ = 0.599 
$\pm$ 0.013 (stat) $\pm$ 0.011 (syst). These results are consistent with the 
previous measurements~\cite{cleo}, $\delta_D$ = $-28^\circ\pm 8^\circ$ (stat) 
$\pm 11^\circ$ (syst) and $r_D$ = 0.52 $\pm$ 0.05 (stat) $\pm$ 0.04 (syst).\\
 The measurement of   $r_D$ and $\delta_D$  is a prerequisite to extract the CKM angle $\gamma$ from the analysis of  $\Bmp \to \Dztilde \Kpm $ decays \cite{ref:GLWADS}, where the  symbol \Dztilde indicates either a \Dz or a \Dzb meson decaying into a CP-eigenstate as $K^- K^+ \piz$ or $\KS \pim \pip$ as we will see in more detail in the next Section. 

\section{Dalitz model and CKM $\gamma$ extraction}

\begin{table*}[!htb]
\caption{Complex amplitudes $a_r e^{i\phi_r}$ and fit fractions of the different components 
($\KS\pim$ and $\KS\pip$ resonances, and $\pip\pim$ poles) obtained from the fit of 
the $\Dz \to \KS\pim\pip$ Dalitz distribution from $\Dstarp \to \Dz \pip$ events. 
Errors are statistical only.
Masses and widths of all resonances are taken from~\cite{PDG_2004}, while the pole masses and scattering 
data are from~\cite{ref:AS}.
The fit fraction is defined for the resonance terms ($\pi\pi$ S-wave term) as the integral 
of $a_r^2 |{\cal A}_r(m^2_-,m^2_+)|^2$  over the Dalitz plane
divided by the integral of $|{\cal A}_D(m^2_-,m^2_+)|^2$. The sum of fit fractions is $1.16$.}
\label{tab:fitreso-likelihood}
\begin{center}
\begin{tabular}{l|ccc}
\hline
    Component  &  $Re \{a_r e^{i\phi_r}\}$ &  $Im \{a_r e^{i\phi_r}\}$ & Fit fraction (\%) \\ 
\hline \hline
$K^{*}(892)^-$        &    $-1.159 \pm 0.022$    &  $1.361 \pm 0.020$   &  $58.9$    \\ 
$K^{*}_0(1430)^-$     &    $2.482 \pm 0.075$     &  $-0.653 \pm 0.073$   &  $9.1$    \\ 
$K^{*}_2(1430)^-$     &    $0.852 \pm 0.042  $   &  $-0.729 \pm 0.051$   &  $3.1$    \\ 
$K^{*}(1410)^-$       &    $-0.402 \pm 0.076$    &  $0.050 \pm 0.072$   &  $0.2$    \\ 
$K^{*}(1680)^-$       &    $-1.00 \pm 0.29$   &  $1.69\pm 0.28$   &  $1.4$    \\ 
\hline
$K^{*}(892)^+$     &    $0.133 \pm 0.008$   &  $-0.132\pm 0.007$  &  $0.7$     \\ 
$K^{*}_0(1430)^+$  &    $0.375 \pm 0.060$   &  $-0.143\pm0.066$  &  $0.2$    \\ 
$K^{*}_2(1430)^+$  &    $0.088 \pm 0.037$   &  $-0.057\pm0.038$  &  $0.0$    \\ 
\hline 
$\rho(770)$         &    1 (fixed)  &    0 (fixed)  &   $22.3$    \\ 
$\omega(782)$       &    $-0.0182\pm 0.0019$   &  $0.0367\pm0.0014$   &     $0.6$     \\ 
$f_2(1270) $        &    $0.787\pm 0.039$    &  $-0.397\pm0.049$   &     $2.7$     \\ 
$\rho(1450)$        &    $0.405\pm 0.079$   &  $-0.458\pm0.116$   &     $0.3$    \\ 
\hline \hline
$\beta_1$           &    $-3.78 \pm 0.13$  &  $1.23\pm0.16$   &     $-$     \\ 
$\beta_2$           &    $9.55 \pm 0.20$   &  $3.43\pm0.40$   &     $-$    \\ 
$\beta_4$           &    $12.97 \pm 0.67$   &  $1.27\pm0.66$   &     $-$    \\ 
$f_{11}^{\rm prod}$ &    $-10.22 \pm 0.32$  &  $-6.35\pm0.39$   &     $-$     \\ 
sum of $\pip\pim$ S-wave & &                 &     $16.2$ \\
\hline
\end{tabular}
\end{center}
\end{table*}

Assuming no \CP asymmetry in $D$ decays  the $\Bmp \to \Dztilde \Kpm $, $\Dztilde \to \KS \pim \pip$,
  decay chain rate $\Gamma_\mp(m^2_-,m^2_+)$ can be \mbox{written as}
\begin{eqnarray}
\Gamma_\mp(m^2_-,m^2_+) \propto |{\cal A}_{D\mp}|^2 + \rb^2 |{\cal A}_{D\pm}|^2 +  \\ \nonumber 
   2 \left\{ \xbmp \re[{\cal A}_{D\mp} {\cal A}_{D\pm}^*]+\ybmp\im[ {\cal A}_{D\mp} {\cal A}^*_{D\pm}] \right\} ~,
\label{eq:ampgen1}
\end{eqnarray}
where $m^2_-$ and $m^2_+$ are the squared invariant masses of the $\KS\pim$ and $\KS\pip$ combinations
respectively from the \Dztilde decay, and ${\cal A}_{D\mp} \equiv {\cal A}_{D}(m^2_\mp,m^2_\pm)$, with
${\cal A}_{D-}$ (${\cal A}_{D+}$) the amplitude of the $\Dz \to \KS\pim\pip$ ($\Dzb \to \KS\pip\pim$) decay.
In Eq.~(11)    the following definitions are used, $\xbmp = \rb \cos(\deltab \mp\gamma)$ and  $\ybmp=\rb \sin(\deltab \mp\gamma)$.  Here,
\rb is the magnitude of the ratio of the amplitudes ${\cal A}(\Bm \to \Dzb \Kstarm)$ and ${\cal A}(\Bm \to \Dz \Kstarm)$  and \deltab is their relative strong phase.

\begin{figure}[!ht]
\begin{center}
\begin{tabular} {cc}  
{\includegraphics[height=4.0cm]{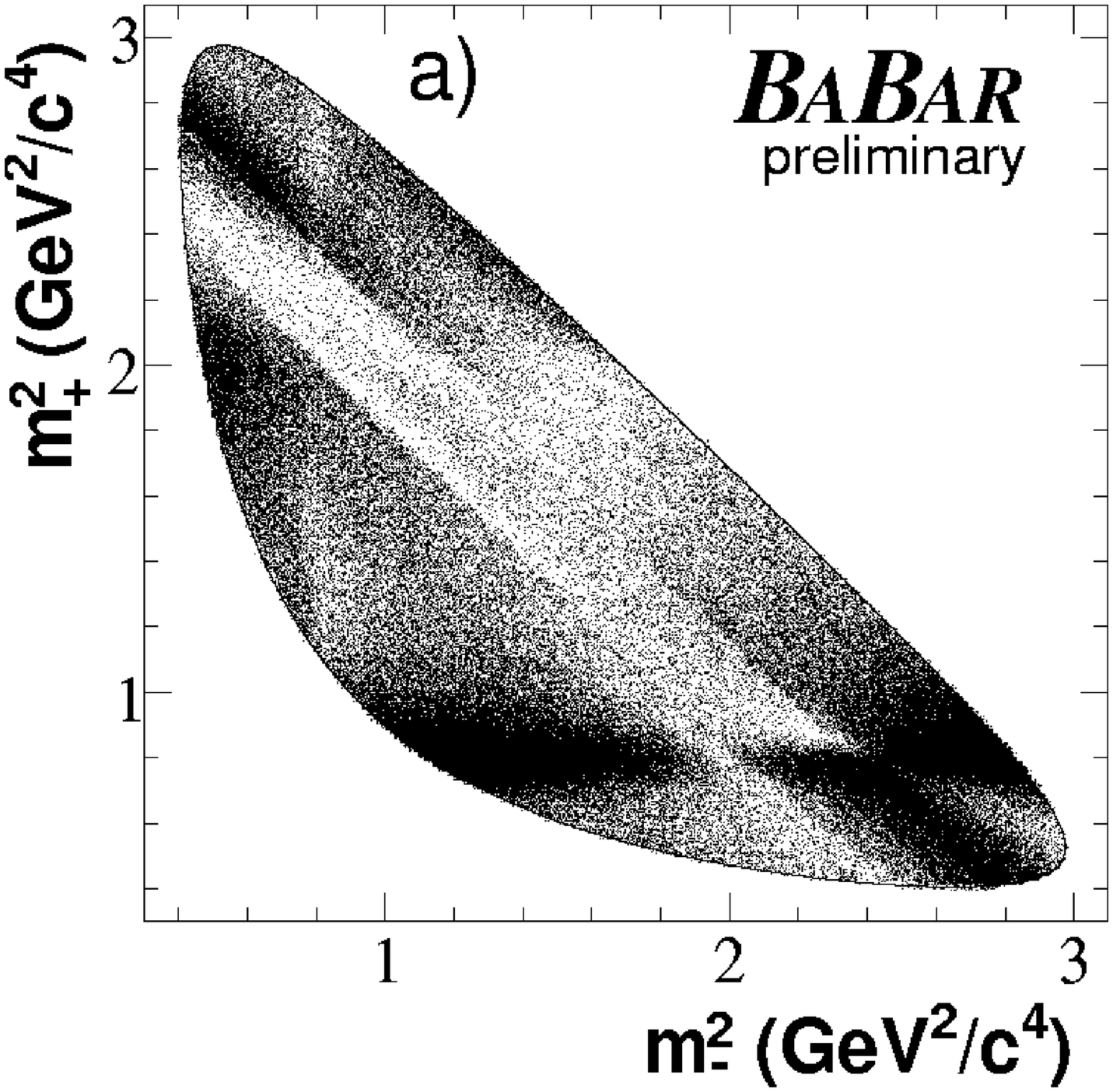}} &
{\includegraphics[height=4.0cm]{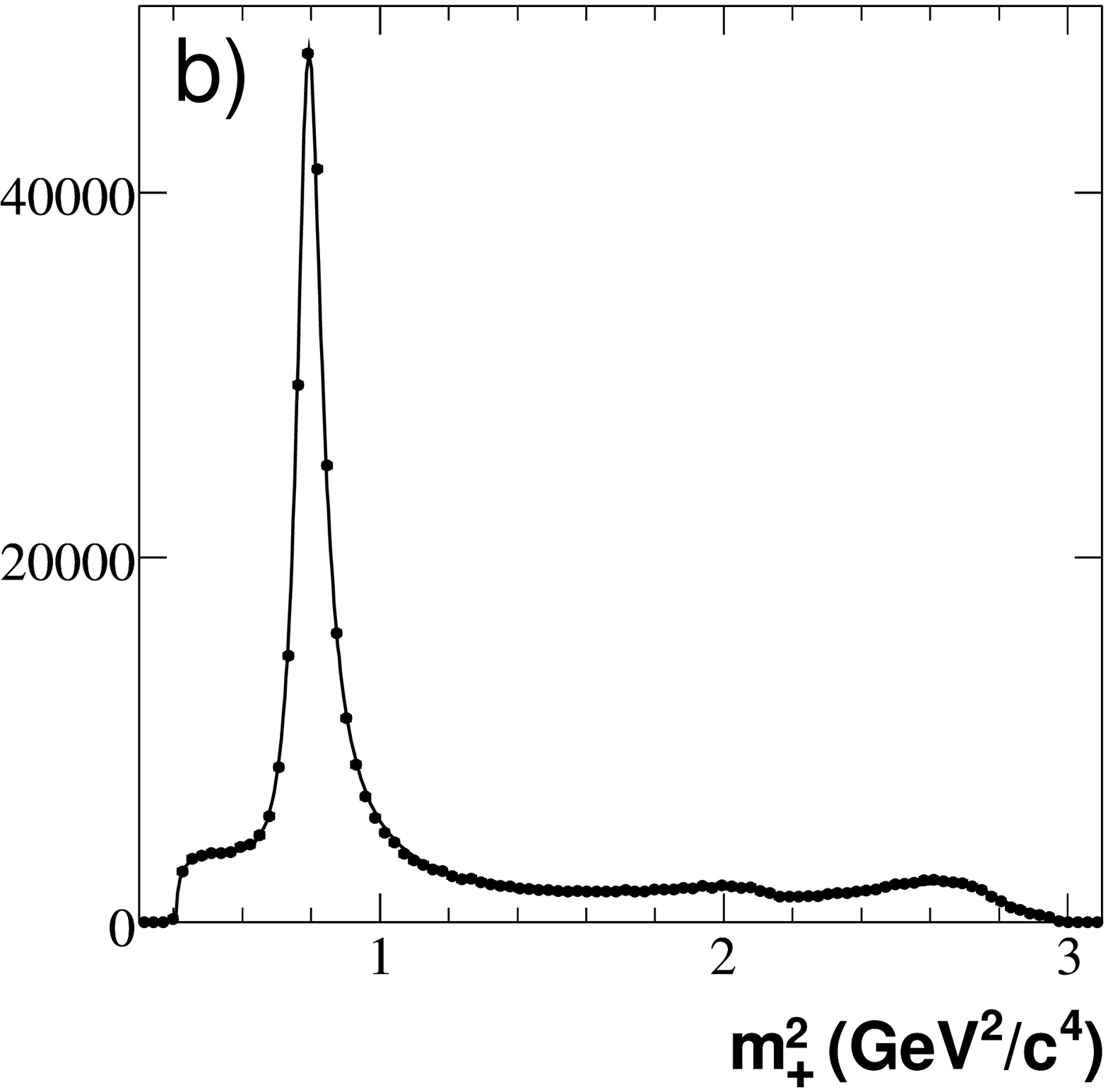}} \\
{\includegraphics[height=4.0cm]{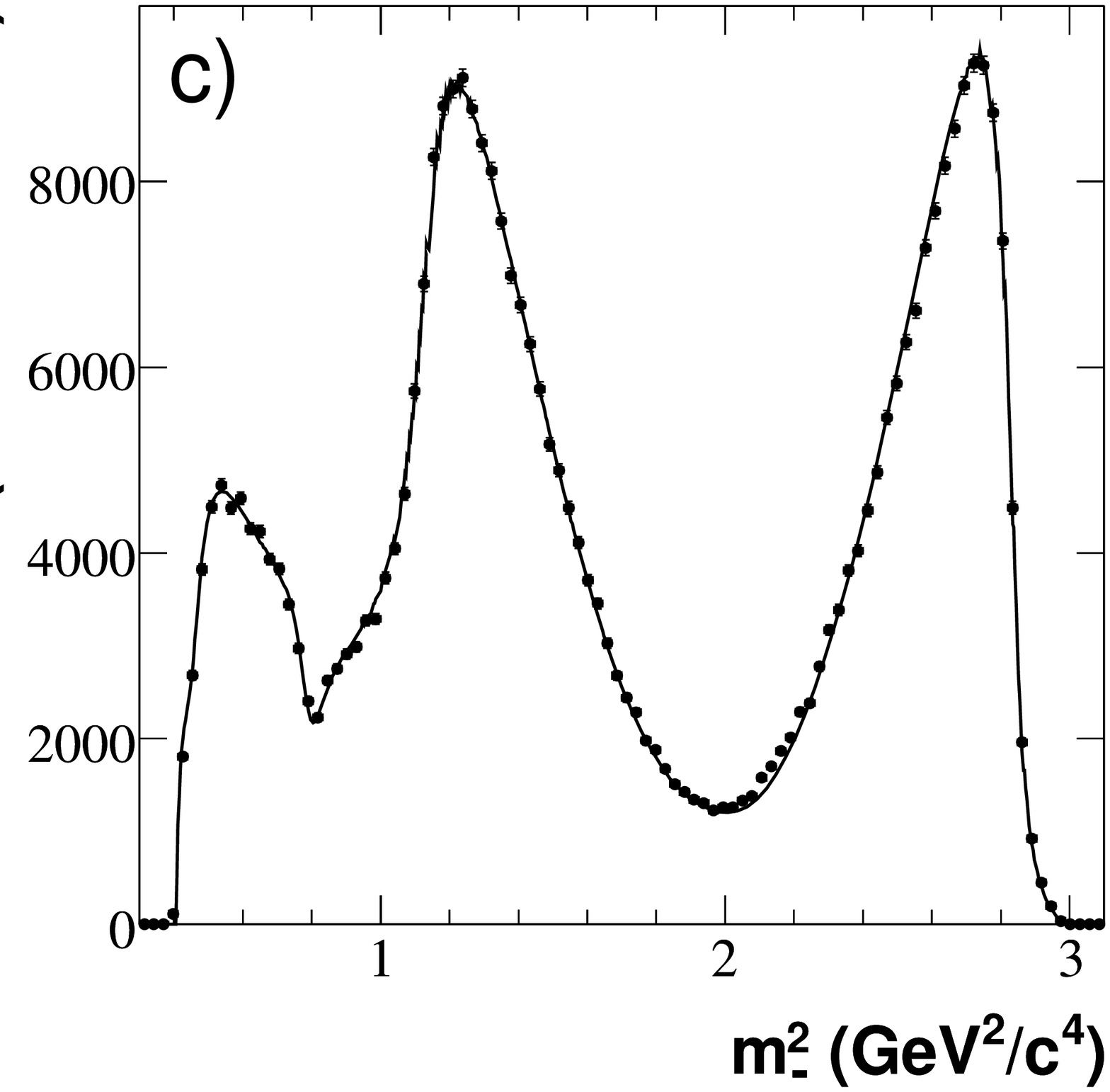}} &
{\includegraphics[height=4.0cm]{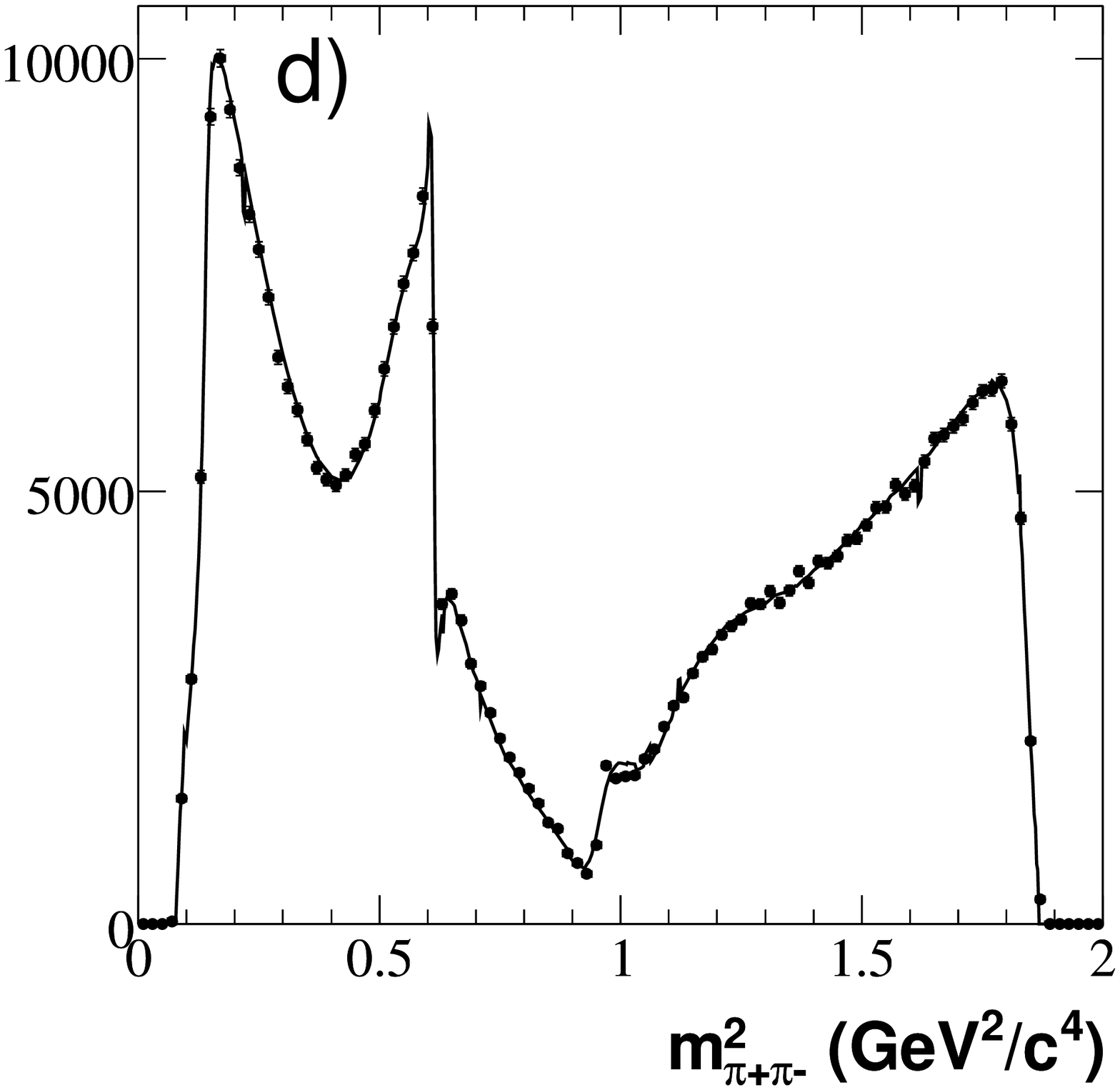}} \\
\end{tabular}   
\caption{(a) The $\bar{D}^0 \to \KS \pim \pip$ Dalitz distribution from $D^{*-} \to \bar{D}^0 \pi^-$ events, and
 projections
on (b) $m^2_+=m^2_{\KS\pi^+}$, (c) $m^2_-=m^2_{\KS\pi^-}$, and (d) $m^2_{\pip\pim}$. $\Dz \to \KS \pi^+ \pi^-$ f
rom $\Dstarp \to \Dz \pip$ events 
are also included. The curves are the reference model fit projections. 
}
\label{fig:BWfit-res}
\end{center}
\end{figure}

Once the decay amplitude ${\cal A}_{D}$ is known, the Dalitz plot distributions for \Dztilde from \Bm and \Bp decays 
can be simultaneously fitted to $\Gamma_-(m^2_-,m^2_+)$ and $\Gamma_+(m^2_-,m^2_+)$ as given by Eq.~(11), 
respectively, and the angle $\gamma$ can be extracted. 

Since the measurement of $\gamma$ arises from the interference term in Eq.~(11), the uncertainty in the 
knowledge of the complex form of ${\cal A}_{D}$ can lead to a systematic uncertainty.

Two different models 
describing the $\Dz \to \KS \pim \pip$ decay have been used in this analysis. The first model (also referred to 
as Breit-Wigner model)~\cite{ref:cleomodel}
expresses ${\cal A}_{D}$ as a sum of 
two-body decay-matrix elements and a non-resonant contribution. In the second model (hereafter referred to as the 
$\pi\pi$ S-wave K-matrix model) the treatment of the $\pi\pi$ S-wave states in $\Dz \to \KS \pim \pip$ uses a K-matrix 
formalism~\cite{ref:Kmatrix,ref:aitchison} to account 
for the non-trivial dynamics due to the presence of broad and overlapping resonances. The two models have been obtained using a high statistics flavor tagged \Dz~ sample ($D^{\ast +} \rightarrow \Dz \pip_s$) selected  from $e^+ e^- \ra c \bar c$ events recorded by BaBar.

 In the Breit-Wigner model a  set of several  two-body amplitudes is used, including  five
Cabibbo-allowed amplitudes: $K^{*}(892)^{+}\pi^-$, $K^{*}(1410)^{+}\pi^-$,
$K^{*}_0(1430)^{+}\pi^-$, $K^{*}_2(1430)^{+}\pi^-$ and $K^*(1680)^+\pi^-$, 
their doubly Cabibbo-suppressed partners, and eight channels with a $K^0_S$ 
and a $\pi\pi$ resonance: $\rho$, $\omega$, $f_0(980)$, $f_2(1270)$, 
$f_0(1370)$, $\rho(1450)$, $\sigma_1$ and $\sigma_2$ . 
The  Breit--Wigner  masses and widths of the scalars
$\sigma_1$ and $\sigma_2$ are left unconstrained, while the parameters of
the other resonances are taken to be the same as in ~\cite{ref:cleomodel}. 
The parameters of the $\sigma$ resonances obtained in the fit are as 
follows: $M_{\sigma_1} = 519 \pm 6$ MeV/$c^2$, $\Gamma_{\sigma_1} = 
454 \pm 12$ MeV/$c^2$, $M_{\sigma_2} = 1050 \pm 8$ MeV/$c^2$ and  
$\Gamma_{\sigma_2} = 101 \pm 7$ MeV/$c^2$ (the errors are statistical only).
 The alternative  model is based on a fit to scattering data (K-matrix \cite{ref:AS}) used to parametrize the $\pi \pi$ S-wave component. This variation  is used to estimate the model systematic uncertainty on $\gamma$ since it gives an equally good fit to data.

The error due to the resonance model can be avoided by using the model-independent 
$\gamma$ measurement proposed in \cite{abi}. In this approach, 
the Dalitz plot is partitioned in bins symmetric  with
respect to the $\pi^+ \pi^-$ axis. Counting the number of events in such bins from entangled $D$ decay samples, in addition to the already
utilized flavour-tagged $D$ decay samples,
 can determine the strong phase variation over the Dalitz plot.
 For this the data of a $\tau$-charm factory is needed. 
Useful samples consist of $\psi(3770) \to D^0 \overline{D}^0$ events 
where one of the $D$ mesons decays into a $CP$ eigenstate (such as $K^+ K^-$ 
or $K_S^0 \omega$), while the $D$ meson going in the opposite direction decays into $K_S^0 \pi^+ \pi^-$.  Using also a similar sample where both mesons from the $\psi(3770)$ decay into the $K^0  \pi^+ \pi^-$ state provides enough information to measure all the needed hadronic parameters in $D$ decay up to one overall discrete ambiguity (this can be resolved using a Breit-Wigner model). CLEO-c showed that  with the current 
integrated luminosity of 280 pb$^{-1}$ at the $\psi(3770)$ resonance,
these samples  are already available.

With the luminosity of 750 pb$^{-1}$, that CLEO-c should get at  the end of its operation,
the samples will be respectively about $1000$ and $2000$ events.
Using these two samples with a binned analysis and assuming $r_B = 0.1$,
a  $4^o$ precision on $\phi_3$  could be  obtained \cite{Bondar:2005ki,Bondar:2007ir}.

\section{Conclusions.}


Charm meson  multi-body decays are crucial  to determine  light strong interaction bound states. The nature of such mesons is still unclear, but more information is emerging from high statistics Dalitz analysis of $D$ decays. In the future multi-channles analyses may be the way to go to identify underline structure of the light mesons. For instance a   measurement of  the couplings  of the S-wave  in various $D_s$ decays can help in interpreting   the   $f_0 (980)$  as two di-quark bound states \cite{multiquark}.
  Determining the decay dynamic of charm mesons is relevant for  method to extract the CKM angle  $\gamma$ in B decays as $B^+  \to D^0 K^+$. The effect of the knowledge of  the strong phase variation in charm meson decay translates into a model systematic error on the $\gamma$ value. Model dependence can be removed if special sample of D meson charm decays in quantum-coherent states will be available, bringing down the model error on $\gamma$  to few degrees.

\begin{acknowledgments}
 The author  warmly thanks  the organizers  for the  great  conference in such a beautiful venue. This work has been supported by the Istituto Nazionale di Fisica Nucleare (INFN), Italy. 

\end{acknowledgments}

\bigskip 

\begin{thebibliography}{99}
\bibitem{chconj}  Charge conjugation is always implied throughout the paper.
\end{thebibliography}

\end{document}